\documentclass{amsart}%%%%%%%%%%%%%%%%%%%%%%%%%%%%%%%%%%%%%%%%%%%%%%%%%%%%%%%%%%%%%%%%%%%%%%%%%%%%%%%%%%%%%%%%%%%%%%%%%%%%%%%%%%%%%%%%%%%%%%%%%%%%%%%%%%%%%%%%%%%%%%%%%%%%%%%%%%%%%%%%%%%%%%%%%%%%%%%%%%%%%%%%%%%%%%%%%%%%%%%%%%%%%%%%%%%%%%%%%%%%%%%%%%%%%%%%%%%%%%%%%%%%%%%%%%%%
\usepackage{amsmath}
\usepackage{amssymb}
\usepackage{amsfonts}

\newtheorem{theorem}{Theorem}

\newtheorem{conclusion}[theorem]{Conclusion}

\newtheorem{corollary}[theorem]{Corollary}

\newtheorem{definition}[theorem]{Definition}

\newtheorem{lemma}[theorem]{Lemma}

\newtheorem{proposition}[theorem]{Proposition}
\newtheorem{remark}[theorem]{Remark}

\newcommand{\cA}{{\mathcal A}}
\newcommand{\cB}{{\mathcal B}}

\newcommand{\cP}{{\mathcal P}}

\newcommand{\cK}{{\mathcal K}}
\newcommand{\cL}{{\mathcal L}}
\newcommand{\cS}{{\mathcal S}}
\newcommand{\cH}{{\mathcal H}}

\newcommand{\bC}{{\mathbb{C}}}
\newcommand{\bN}{{\mathbb{N}}}
\newcommand{\Cs}{{{$\hbox{\bf C}^*$}}}
\DeclareMathOperator{\Exp}{Exp}
\DeclareMathOperator{\Ext}{Ext}
\DeclareMathOperator{\Tr}{Tr}
\dedicatory{To SLAVA} 

\begin{document}

\title{On positive maps in quantum information}

\author{W{\l}adys{\l}aw A. Majewski}
\address{Institute of Theoretical Physics and Astrophysics, Gda{\'n}sk
University, Wita Stwo\-sza~57, 80-952 Gda{\'n}sk, Poland and Unit for BMI, North-West-University, Potchefstroom, South Africa} 
\email{fizwam@univ.gda.pl}

\begin{abstract}
Using Grothendieck approach to the tensor product of locally convex spaces we review a characterization of positive maps as well as Belavkin-Ohya characterization of PPT states. Moreover, within this scheme, \textit{ a generalization of the idea of Choi matrices for genuine quantum systems will be presented}.
\end{abstract}
\maketitle
\section{Introduction}
In Quantum Physics, Operator Theory, and recently in Quantum Information some of the most important families of linear maps are positive, k-positive, completely positive, and decomposable. This motivates various attempts to understand and describe the structure of the family of positive maps as well as some of its subsets.

In this notes we wish to review and elaborate some Slava Belavkin results related to the theory of positive maps.
To be more concrete, we will be concerned with the following two questions: Is there an approach allowing a generalization, for infinite dimensional case,  the present characterization of positive maps? This question is crucial for that characterization of quantum maps which is based on the concept of Choi matrices. We recall that the idea of Choi matrices is the starting point of many papers devoted to an analysis of quantum maps.

In the second part of these notes we will be addressed to the second question, i.e. to a characterization of the so called family of PPT states.
Both problems will be treated within the scheme given by the theory of tensor products of Banach spaces.

The paper is organized as follows. In the next Section, we will provide the notation, preliminaries,  some basic result from the theory of positive maps, and the Grothendieck theory of tensor product of Banach spaces. Section 3 will be devoted to a review of Slava's results on Radon-Nikodym derivatives in the context of recent characterization of quantum maps. Moreover, a generalization of Choi matrices will be presented. In Section 4, an analysis of PPT states will be given, and again, this will be done for a general case.

\section{Definitions and notations}

As it was mentioned, in this section we provide a summary of some basic facts on the theory of positive maps on the ordered structures with an emphasis on convex structures.
To begin with, let $\mathfrak{A}$ and $\mathfrak{B}$ be \Cs-algebras (with a unit), $\mathfrak{A}_h = \{ a \in \mathfrak{A}:
a = a^* \}$ -- the set of all self adjoint elements in $\mathfrak{A}$, $\mathfrak{A}^+ = \{ a \in \mathfrak{A}_h: a \ge 0 \}$ -- the set of all positive elements in 
$\mathfrak{A}$, and $\cS(\mathfrak{A})$ the set of all states on $\mathfrak{A}$, i.e. the set of all linear functionals $\varphi$
on $\mathfrak{A}$ such that $\varphi(1) = 1$ and $\varphi(a)\geq0$ for any $a \in \mathfrak{A}^+$.
In particular
$$ (\mathfrak{A}_h, \mathfrak{A}^+)\text{ is an ordered  Banach  space.} $$
We say that a linear map $\alpha : \mathfrak{A} \to \mathfrak{B}$ is positive if $\alpha(\mathfrak{A}^+) \subset \mathfrak{B}^+$.
The set of all (linear, bounded) positive maps $\alpha: \mathfrak{A} \to \mathfrak{B}$ will be denoted by $\cL^+(\mathfrak{A}, \mathfrak{B})$.
 We say that $\alpha \in \cL^+(\mathfrak{A}, \mathfrak{B})$ is an extreme point of $\cL^+(\mathfrak{A}, \mathfrak{B})$
if there are no two different maps $\alpha_1$ and $\alpha_2$ in $\cL^+(\mathfrak{A}, \mathfrak{B})$ such that 
$\alpha = \lambda \alpha_1 + (1 - \lambda) \alpha_2$ with $\lambda \in (0,1)$.
The set of extreme points of $\cL^+(\mathfrak{A}, \mathfrak{B})$ will be denoted by $\operatorname{Ext} \{\cL^+(\mathfrak{A}, \mathfrak{B})\}$.
We recall (Krein-Milman theorem) that a compact convex set $C$ is a (closed) convex hull of its extreme points. Hence, the subset $\cL^+_1(\mathfrak{A}, \mathfrak{B})$ of unital maps in $\cL^+(\mathfrak{A}, \mathfrak{B})$ can be described in terms of extreme unital positive maps.

Consequently, it was tempting to characterize the set of positive maps by means of their extremal points.
To this end, one should
find all extremal maps in $\cL^+(\mathfrak{A}, \mathfrak{B})$. In the very special case, for unital maps if $\mathfrak{A} \equiv M_2(\bC)$
and $\mathfrak{B} \equiv M_2(\bC)$, the extremal maps were described by St{\o}rmer \cite{StActa}. Some partial results in this direction, for the case $M_2(\bC) \to M_{n+1}(\bC)$ with $n\geq 2$, were obtained in \cite{MajMar} while more generally, positive maps on low dimensional matrix algebras were studied in \cite{W}. However, these and other cases are so complicated that it seems that the program of finding all extremal positive maps is too difficult. Therefore, one can turn to a special subset of extremal positive maps.

\begin{definition}
Let $C$ be a convex set in a Banach space $X$. A point $x \in C$ is an exposed point of $C$ ($x \in \Exp\{C\}$) 
if there is $f \in X^*$ (dual of $X$) such that $f$ attains its maximum on $C$ at $x$ and only at $x$.
\end{definition}

In general, one has $\Ext\{C\} \supseteq \Exp\{C\}$ but even in low dimensional case, there are simple examples of 2-dimensional convex compact sets such that the inclusion $\Ext\{C\} \supset \Exp\{C\}$ is proper (see \cite{FLP}). On the other hand, as behind this definition is a variational principle, the nature of exposed points looks more operationally efficient than that of extreme points.

Our interest in exposed points stems from the following result (see \cite{Strasz}, \cite{Klee} and \cite{FLP})

\begin{proposition}
Every norm-compact convex set $C$ in a Banach space $X$ is the closed convex hull of its exposed points.
\end{proposition}

As the assumption of norm-compactness seems to be rather strong we note that there are several extensions of this proposition, eg to locally compact sets (see \cite{FLP}), 
to weakly compact sets (see \cite{Lin} and references given there).

Now we wish to provide some preliminaries from the theory of tensor products of Banach spaces.
Let $X$, $Y$ be Banach algebras. 
We denote by $X \odot Y$ the algebraic tensor product of $X$ and $Y$
 (algebraic tensor product of two $^*$-Banach algebras is defined as tensor product of two vector spaces with $^*$-algebraic structure determined by the two factors; so the topological questions are not considered). 
We consider two norms on $X \odot Y$ (cf \cite{Gro}). Firstly, the projective norm:
\begin{equation}
\label{projektywna norma}
||u||_{\pi} = \inf \{ \sum_{i=1}^{n}\Vert x_i \Vert \Vert y_i\Vert: \quad u = \sum_{i=1}^n x_i \otimes y_i, \quad x_i \in X, \quad y_i \in Y\}.
\end{equation}

We denote by $X \otimes_{\pi}Y$ the completion of $X \odot Y$ with respect to the projective norm $||\cdot||_{\pi}$ and this Banach space will be referred to as the projective tensor product of the Banach spaces $X$ and $Y$.

Secondly, the injective norm is defined as
\begin{equation}
\label{injektywna norma}
||u||_{\epsilon} = \sup \{ | \sum_{i=1}^{n}f(x_i)g(y_i)|:  f\in X^*, \Vert f \Vert \leq1; \quad g \in Y^*, \Vert g \Vert \leq 1 \}
\end{equation}
where $ u = \sum_{i=1}^n x_i \otimes y_i, \quad x_i \in X, \quad y_i \in Y$.

We denote by $X \otimes_{\epsilon}Y$ the completion of $X \odot Y$ with respect to the injective norm $||\cdot||_{\epsilon}$ and this Banach space will be referred to as the injective tensor product of the Banach spaces $X$ and $Y$.

Denote by $\mathfrak{B}(X \times Y)$ the Banach space of bounded bilinear mappings $B$ from $X \times Y$ into the field of scalars with the norm given by $||B|| = \sup \{ |B(x,y)|; \Vert x \Vert \leq 1, \Vert y \Vert \leq 1 \}$.
Note (for all details see \cite{Ryan}, \cite{DF}, \cite{Diestel}), that with each bounded bilinear form $B \in \mathfrak{B}(X \times Y)$ there is an associated operator $L_B \in \cL(X, Y^*)$ defined by $\langle y, L_B(x)\rangle = B(x,y)$.  The mapping $B \mapsto L_B$ is an isometric isomorphism between the spaces $\mathfrak{B}(X \times Y)$ and $\cL(X, Y^*)$. Hence, there are identifications
\begin{equation}
\label{lala}
\mathfrak{B}(X \times Y) = \cL(X, Y^*),
\end{equation}
and (cf \cite{Ryan})
\begin{equation}
\label{lala1}
(X \otimes_{\pi} Y)^* =  \cL(X, Y^*),
\end{equation}
if the action of operator $L : X \to Y^*$ is considered also as a functional
\begin{equation}
<\sum_{i=1}^n x_i \otimes y_i, L> = \sum_{i=1}^n <y_i, Lx_i>.
\end{equation}
Thus, the dual $(X \otimes_{\pi} Y)^*$ can be identified with the space of continuous bilinear forms. There is a natural question: which bilinear forms $\zeta$ are continuous with respect to $\Vert \cdot \Vert_{\epsilon}$?
The answer is that a bounded linear functional $\zeta$ on $X\otimes_{\epsilon}Y$ is uniquely determined by a regular Borel measure $\mu$ on the compact space $B_{X^*} \times B_{Y^*}$, $B_{Z^*}$ stands for a unit ball in $Z^*$, in such a way that
\begin{equation}
\label{lala2}
\zeta(x,y) = \int_{B_{X^*}\times B_{Y^*}} \psi(x)\phi(y) d\mu(\psi,\phi) .
\end{equation}

Such forms are called integral, and the corresponding operators are called integral. 
The set of them will be denoted by $\mathfrak{B}_I(X \times Y)$ and $\mathfrak{I}(X, Y^*)$ respectively. Moreover, one has
\begin{equation}
(X\otimes_{\epsilon}Y)^* \subseteq (X \otimes_{\pi}Y)^* = \cL(X, Y^*).
\end{equation}

and
\begin{equation}
\label{lalala}
(X\otimes_{\epsilon}Y)^* = \mathfrak{B}_I(X \times Y) = \mathfrak{I}(X, Y^*).
\end{equation}

We end this brief review of tensor products with the following observation (for details see \cite{Jarch}). Assume that a Banach space $X$ has its predual, i.e. there is a Banach space $X_*$ such that $(X_*)^* \cong X$. Then the canonical bilinear form $<\cdot, \cdot> : X \times X_* \to \bC$ is continuous. Hence, its linearization
\begin{equation}
tr: X \otimes_{\pi} X \to \bC
\end{equation}
is continuous and it is called the trace functional on $X \otimes_{\pi} X$. Moreover for $u = \sum_{i=1}^n x_i \otimes y_i \in X \odot X_*$ the number
\begin{equation}
\label{5bis}
tr(u) = \sum_{i=1}^n <x_i,y_i>
\end{equation}
is called the trace of $u$, and it does not depend on the particular representation of $u$.
Moreover, it is worth pointing out that although $tr$ is continuous for the topology given by $\|\cdot \|_{\pi}$, in general, it is not continuous for the topology given by $\| \cdot \|_{\epsilon}$
as, in general, the identity operator is not the integral one (see \cite{Jarch} for details).

To use the above outlined theory of tensor products of Banach spaces to a description of positive maps, let $\cB(\cK)$ be the algebra of all linear, bounded operators on a Hilbert space $\cK$.
 ${\mathfrak T} \equiv {\mathfrak T}(\cH)$ will denote the set of trace class operators on a Hilbert space $\cH$. $\cB(\cH) \ni x \mapsto x^t \in \cB(\cH)$ stands for the transpose map of $\cB(\cH)$ with respect to some orthonormal basis.
The set of all linear bounded (positive) maps $\phi: \cB(\cK) \to \cB(\cH)$ will be denoted by $\cL(\cB(\cK), \cB(\cH))$ ($\cL^+(\cB(\cK), \cB(\cH))$ respectively). Then

\begin{lemma} (Basic)
\label{pierwszylemat}
There is an isometric isomorphism $\phi \mapsto \tilde{\phi}$ between $\cL(\cB(\cK), \cB(\cH))$ and
$(\cB(\cK) {\otimes}_{\pi} {\mathfrak T})^*$ given by
\begin{equation}
\label{5}
(\tilde{\phi})(\sum_{i=1}^n a_i\otimes b_i) = \sum_{i=1}^n \Tr(\phi(a_i)b^t_i),
\end{equation}
where $\sum_{i=1}^n a_i\otimes b_i \in \cB(\cK)\odot {\mathfrak T}$.

Furthermore, $ \phi \in \cL^+(\cB(\cK), \cB(\cH))$ if and only if $\tilde{\phi}$ is positive on $\cB(\cK)^+ {\otimes}_{\pi} {\mathfrak T}^+$.
\end{lemma}

To comment on this result we make
\begin{remark}
\begin{enumerate}
\item $\mathfrak T$ appears in Lemma \ref{pierwszylemat} as ${\mathfrak T}^* = \cB(\cH)$ (so in formula (\ref{lala}), one puts $Y = \mathfrak T$, obviously $X$ is taken to be $\cB(\cK)$).
\item To understand the right hand side of (\ref{5})  we note that the canonical bilinear form on $\cB(\cH) \times \cB(\cH)_*$ is given by the canonical trace $\Tr$ on $\cB(\cH)$ and apply (\ref{5bis}).
\item The formula (\ref{5}) does not take into account weak topologies. On the other hand, to speak about weak$^*$ continuity of $\tilde{\phi}$, such topologies are crucial. We will come to this point in the next Section.
\item There is no restriction on the dimension of Hilbert spaces. In other words, this result can be applied to the true quantum systems.
\item In \cite{St2}, see also Chapter 4 in \cite{Stbook},  St{\o}rmer showed that in the special case when ${\mathfrak A} = M_n(\bC)$ and $\cH$ has dimension equal to $n$, the above Lemma is a reformulation of Choi result \cite{Ch1}, \cite{Ch3}.
\item  One should note that the positivity  of the functional  $\tilde{\phi}$ is defined by the cone
${\mathfrak A}^+ {\otimes}_{\pi} {\mathfrak T}^+$ (for another definitions of positivity in tensor products see \cite{Witt}, and \cite{MMO}). In particular, positivity determined by $({\mathfrak A} \otimes_{\pi} {\mathfrak T})^+$ leads to completely positive maps. This remark will be employed in the last Section.
\item Some forms of Basic Lemma, being a consequence of Grothendieck's approach, are known at least from late sixties, see \cite{WW} and references given there. The form of Lemma \ref{pierwszylemat} is taken from \cite{St85}.
\end{enumerate}
\end{remark}

\vskip 0.5cm

We wish to close this section with a basic package of terminology used in operator algebras.
Let us recall that for a Hilbert space $\cL$
every normal (so $^*$-weakly continuous) state $\phi$ on $\cB(\cL)$ has the form of
$\phi(a)=\Tr(\varrho a)$, where $\varrho$ is a uniquely determined
\textit{density matrix},
i.e. an element of $\cB(\cL)^+$ such that $\Tr\varrho=1$.

A linear map $\alpha : \cB(\cH) \to \cB(\cK)$ is called $k$-positive if a map $id_{M_k} \otimes \alpha : M_k \otimes \cB(\cH) \to M_k \otimes \cB(\cK)$ is positive, where $M_k \equiv M_k(\bC)$ denotes the algebra of $k\times k$ matrices with complex entries. A map $\alpha$ is called completely positive if it is $k$-positive for any $k$. A completely positive map $\alpha$ will be shortly called a CP map.
A positive map $\alpha:\cB(\cH) \longrightarrow\cB(\cK)$ is called
\textit{decomposable}
if
there are completely positive maps $\alpha_1,\alpha_2:\cB(\cH)\longrightarrow\cB(\cK)$
such that $\alpha=\alpha_1+\alpha_2\circ\tau_\cH$, where $\tau_\cH$ stands for a transposition map on $\cB(\cH)$. Let $\cP$, $\cP_c$ and $\cP_d$
denote the set of all positive, completely positive
and decomposable maps from $\cB(\cH)$ to $\cB(\cK)$, respectively. Note that
$$ %\be
\cP_c \subset \cP_d \subset \cP
$$ %\ee{inclP}
(see also \cite{Ch1}- \cite{Ch2}).

A state $\varphi$ on $\mathfrak{A} \otimes \mathfrak{B}$
is said to be \textit{separable} if it can be approximated (in norm) by states of the 
form
$$
\varphi=\sum_{n=1}^N p_n\varphi_n^1\otimes\varphi_n^2
$$ %\ee{sep}
where $N\in\bN$, $\varphi_n^1$ is a state on $\mathfrak{A}$, $\varphi_n^2$ is a state on $\mathfrak{B}$
for $n=1,2,\ldots,N$, $p_n$ are positive numbers such that
$\sum_{n=1}^Np_n=1$. The set of
all separable states on the algebra
$\mathfrak{A} \otimes \mathfrak{B}$ is denoted by $\cS_s$. A state which is not
in $\cS_s$ is called \textit{entangled} or \textit{non-separable}.

Finally, let us define the family of \textit{PPT} (transposable) states on
$\cB(\cH) \otimes \cB(\cK)$
$$ %\be
\cS_t=\{\varphi\in\cS:\mbox{$\varphi\circ(id_{\cB(\cH)} \otimes \tau_\cK)\in\cS$}\}.
$$ %\ee{St}
where, as before, $\tau_{\cK}$ stands for the transposition map, now defined on $\cB(\cK)$.
Note that due to the positivity of the transposition $\tau_\cK$ every
separable state $\varphi$ is transposable, so
$$ %\be
\cS_s\subset\cS_t\subset\cS.
$$ %\ee{inclS}

We close this Section with a remark that PPT states are of fundamental importance for Quantum Computing. Furthermore, 
it is worth pointing out that these states are closely related to another important concept - to quantum correlations (entanglement) (cf \cite{MajOSID}, \cite{I}).

\section{Positive maps, Radon-Nikodym derivatives, generalized Choi matrices}

Basic Lemma, being a result of metric theory of tensor products, provides the isomorphism
$\phi \mapsto \tilde{\phi}$ between
 $\cL(\cB(\cH), \cB(\cH))\equiv\cL( \cB(\cH))$ and $(\cB(\cH) \otimes_{\pi} \cB(\cH))^*$.
It has a property that $\phi \in \cL^+(\cB(\cH))$ if and only if $\tilde{\phi}$ is positive on $\cB(\cH)^+ \otimes_{\pi} \cB(\cH)^+$.
Further, note that identifying the real algebraic tensor product $\cB(\cH)_h \odot \cB(\cH)_h$ of self-adjoint parts of $\cB(\cH)$ with a real subspace of $\cB(\cH) \odot \cB(\cH)$, one has $\cB(\cH)_h \odot \cB(\cH)_h = (\cB(\cH) \odot \cB(\cH))_h$. Obviously, this can be extended for the corresponding closures.  From now on, we will use these identifications and we will study certain subsets of real tensor product spaces.

The next easy observation says that the discussed isomorphism sends the set $\Exp \{ \cL^+(\cB(\cH)) \}$ onto the set $\Exp \{(\cB(\cH) \otimes_{\pi} \cB(\cH))^{*,+} \} $, where $(\cB(\cH) \otimes_{\pi} \cB(\cH))^{*,+}$ stands for functionals 
on $\cB(\cH) \otimes_{\pi} \cB(\cH)$
which are positive on $\cB(\cH)^+ \otimes_{\pi} \cB(\cH))^+$.

This observation was the starting point of a characterization of positive maps given in \cite{JMP}. Here we briefly summarize this characterization. Assume, for a moment, that underlying Hilbert spaces are finite dimensional.
In that way, one can restrict oneself, to norm-topologies only. In particular, we can apply the basic lemma without any modifications.
 Hence, see \cite{JMP}, any (linear, bounded) functional in $( \cB(\cH) \otimes_{\pi} \cB(\cH))^{*,+}$ is of the form
\begin{equation}
\label{5a}
\varphi(x \otimes y) = \Tr \varrho_{\varphi} \ x \otimes y,
\end{equation}
 with $\varrho_{\varphi}$ being a ``density'' matrix satisfying the following positivity condition (frequently called ``block-positivity'', and denoted ``bp'' for short)
\begin{equation}
\varrho_{\varphi} \geq_{bp}0 \quad  \Leftrightarrow  \quad (f \otimes g, \varrho_{\varphi} f \otimes g) \geq 0,
\end{equation}

for any $f,g \in \cH$.
This ``density matrix'' can be identified with the Choi matrix (cf Remark 4(4)).
To take into account that the isomorphism given in Lemma 3 is also isometric, note that  $\cL(\cB(\cH))$ is equipped with the Banach space operator norm $\Vert \cdot \Vert$. 
Define, see \cite{JMP},

\begin{definition}
The dual norm $\alpha$ to the projective norm $\| \cdot \|_{\pi}$ is defined as
\begin{equation}
\alpha(\varrho_{\varphi}) = \sup_{0\neq a \in 
\cB(\cH) \otimes_{\pi} \cB(\cH)} \frac{|\Tr \varrho_{\varphi} a|}{\| a \|_{\pi}}.
\end{equation}
\end{definition}

An application of Proposition IV.2.2 in \cite{Tak} (see also \cite{Schat}) shows that $\alpha(\cdot)$ is well defined cross-norm.
It will be useful to note that for bp density matrix $\varrho_{\varphi}$
\begin{equation}
\alpha(\varrho_{\varphi}) \geq \frac{|\Tr \varrho_{\varphi}|}{\pi(1)} \quad \text{and} \quad  |\Tr\varrho_{\varphi}| = \Tr \varrho_{\varphi}\leq n \alpha(\varrho_{\varphi}),
\end{equation}
where we have used that $\| 1 \|_{\pi} = n$ and denoted by $1$ the unit of the algebra.

Define
\begin{equation}
\label{Do}
{\mathfrak D}_0 = \{ \varrho_{\varphi}: \alpha(\varrho_{\varphi}) = 1, \  \varrho_{\varphi}= \varrho_{\varphi}^*, \ \varrho_{\varphi}\geq_{bp} 0 \}.
\end{equation}

Basic Lemma and the above discussion say that there is an isometric isomorphism between the set of positive linear maps in  $\cL(\cB(\cH))^+$ of norm one and ${\mathfrak D}_0$. Moreover, for each $\varrho_{\varphi} \in {\mathfrak D}_0$ one has $\Tr \varrho_{\varphi} \leq n \alpha(\varrho_{\varphi}) = n$.

To proceed (cf \cite{JMP}) with the analysis of Basic Lemma, we note that the formula (\ref{5}) 
says that any bp density matrix $\varrho_{\phi}$ determined by an unital positive map $\phi$ has the following normalization

\begin{equation}
\label{8}
\Tr \varrho_{\phi} \equiv \Tr_{\cH \otimes \cH} \varrho_{\phi} = \Tr_{\cH} \phi(1) = n. 
\end{equation}

Conversely, assume that $\varrho_{\phi} \in {\mathfrak D}_0$ and $\Tr \varrho_{\phi} = n$. Then, there exists a linear positive bounded map $\phi$ (of norm one) such that  $\Tr \phi(1) = n$.

However,
\begin{equation}
\Vert \phi(1) \Vert \leq \Vert \phi \Vert \Vert 1 \Vert = \alpha(\varrho_{\phi}) \cdot 1 = 1. 
\end{equation}
Thus $\Vert \phi(1)\Vert \leq 1$. But $\phi(1) \geq 0$ has the following spectral decomposition
$$ \phi(1) = \sum_i \lambda_i E_i ,$$
with $\lambda_i \leq 1$ for all i, and $\sum_i \lambda_i = n$. This means that $\phi(1) = 1$.

To sum up, firstly, we give (see \cite{JMP})

\begin{definition}
The set of bp normalized density matrices is defined as
\begin{equation}
\mathfrak{D} = \{ \varrho_{\phi}: \alpha(\varrho_{\phi}) = 1, \  \varrho_{\phi}= \varrho_{\phi}^*, \ \varrho_{\phi}\geq_{bp} 0, \ \Tr\varrho_{\phi} = n \}.
\end{equation}
\end{definition}

Then, denoting by $\mathfrak C$ the set of unital maps in $\cL(\cB(\cH))^+$, we have, see \cite{JMP} 
\begin{theorem}
\begin{enumerate}
 \item
 Lemma \ref{pierwszylemat} gives an isometric isomorphism between the convex set of unital positive maps $\mathfrak{C}$ and the subset $\mathfrak{D}$ 
of bp normalized density matrices. This isomorphism sends exposed points of the set of all positive unital maps onto exposed points of $\mathfrak{D}$.
\item $\mathfrak D$ is a convex compact set and ${\mathfrak D} = \overline{conv}(Exp({\mathfrak D}))$.
\end{enumerate}
\end{theorem}

The detailed characterization of the set $\mathfrak{D}$ was done in \cite{JMP}, see also \cite{JMP1}. 

\bigskip

However, Quantum Mechanics as well as many problems in quantum information theory need infinite-dimensional Hilbert spaces. Namely, see \cite{Win}, \cite{Wiel}, 
\begin{proposition}
It is impossible to find two elements  $a$, $b$ in a Banach algebra $\cA$ such that 
\begin{equation}
ab - ba = 1.
\end{equation}
\end{proposition}

In other words, it is impossible to carry out a canonical quantization on finite dimensional spaces.
Consequently, \textbf{ it is difficult to speak about quantumness of finite dimensional systems}.
Therefore, we drop the assumption that Hilbert spaces are finite dimensional and \textit{ we will continue our analysis for the general case}.
In particular, \textit{relations established in Lemma \ref{pierwszylemat} should be complemented with arguments taking into account weak topologies}. Furthermore,
it is worth pointing out that such approach involves a generalization of the one-to-one correspondence  between block-positive (positive) operators on $\bC^N \otimes \bC^M$ and positive (CP, respectively) maps $T: M_n \to M_m$
(known as the ``Jamiolkowski isomorphism'' in the quantum information community or as the Choi matrix concept in the mathematical community (cf Remark 4(5)).

\smallskip

To illustrate the necessity of such modifications we recall one of Bielavkin's result. Namely, we  consider a  characterization of quantum channels, which are described by CP maps. Since the underlying Hilbert space $\cH$ 
should be infinite dimensional one, Choi matrices (without any generalization) are not able to describe genuine quantum channels.
Therefore, to compare quantum channels 
and to avoid an employment of Choi matrices
Slava has proposed to use a version of Radon-Nikodym theorem for CP maps.
We cite his main result (see \cite{BS}):

\begin{definition}
Let $\phi, \psi \in CP(\mathfrak{ A}, \cB(\cH)).$ $\phi$ is said to be completely absolutely continuous with respect to $\psi$ if for any $n\in \bN$
$$ \inf_m \sum_{ik=1}^n (f_i, \psi(a^*_{im}a_{km} f_k) = 0,$$
where $f_i\in \cH$, for any increasing family of matrices $\{ [a^*_{im}a_{km}] \}$ implies
$$\inf_m \sum_{i,k=1}^n (f_i, \phi(a^*_{im} a_{km})f_k) = 0,$$
for any $f_k \in \cH$, $j=1,2,...n.$
\end{definition}

Here $CP({\mathfrak A}, \cB(\cH))$ stands for the set of CP maps $\alpha: {\mathfrak A} \to \cB(\cH)$, where $\mathfrak A$ is a $C^*$-algebra. Slava has proved the following important result (see \cite{BS}, \cite{B}):

\begin{theorem}
Let $\phi, \psi \in CP(\mathfrak{A}, \cB(\cH)),$ and let
$$\psi(a) = F^* \pi(a)F $$
be the Stinespring representation for $\psi$, where $F$ is assumed to be bounded operator $\cH \to \cH_0$. Then
$\phi$ is completely absolutely continuous with respect to $\psi$ if and only if it has a spacial representation $\phi(a) = K^* \pi(a)K$ with $\pi(a) K = W \pi(a) F$, where $W$ is a densely defined operator in the minimal $\cH_0$, commuting with $\pi(\cA)= \{ \pi(a), a \in \cA \}$ on the linear manifold $\mathfrak{M} = \{\sum_j \pi(a_j)F f_j \}.$
\end{theorem}
 Consequently, one can compare certain quantum channels for genuine quantum systems. It should be noted that similar Radon-Nikodym type theorem was obtained independently by Arveson (under different assumptions, see \cite{A}).

\smallskip

However, these results do not provide a characterization of quantum operations (see \cite{Raginsky}) and this task needs a generalization of Choi matrices. Here we wish to present such a generalization. We emphasize that such generalization enables us an extension of the given characterization of unital positive maps for quantum systems. 

\begin{remark}
The extension of Choi-Jamiolkowski correspondence for infinite dimensional systems was recently studied by Holevo (see \cite{H}, \cite{H1}).
His approach is based on an analysis of certain positive sesqulinear form $\Omega_{\phi}$ defined on a subset of the tensor product of two (infinite dimensional) Hilbert spaces.
On the other hand, the presented approach is stemming from the theory of tensor product of locally convex vector spaces. Consequently, these two approaches are different.
\end{remark}

As the first step of generalization of Choi-Jamiolkowski correspondence we will discuss direct consequences of Lemma \ref{pierwszylemat}, i.e. the results based on the projective tensor product of $(\cB(\cH), \| \cdot \|)$
and $(\cB(\cH)_*, \| \cdot \|_{tr})$, where $\| \cdot \|_{tr}$ stands for the trace norm.
To this end we will employ the metric theory of tensor products which was outlined in Section II.
Denote by ${\mathfrak K} \equiv  {\mathfrak K}(\cK)$ the $C^*$ algebra of compact operators on a Hilbert space $\cK$. Finally, cf Section 2, $\mathfrak{ T} \equiv {\mathfrak T}(\cK)$ is the Banach algebra of trace class operators on a Hilbert space $\cK$.

 Firstly note
\begin{equation}
\cB(\cH) \odot {\mathfrak T} \subseteq \cB(\cH) \odot {\mathfrak K}
\end{equation}
and
\begin{equation}
closure^{\|.\|}\{\cB(\cH) \odot {\mathfrak T} \} = \cB(\cH) \otimes {\mathfrak K}
\end{equation}
where the closure is taken with respect to the operator norm, $\otimes$ denotes the complete tensor product with respect to the operator norm of two $C^*$ algebras $\cB(\cH)$ and $\mathfrak K$.

Recall that the projective norm 
$\| \cdot \|_{\pi}$ is the largest cross-norm while the injective norm $\| \cdot \|_{\epsilon}$ is the smallest cross-norm. This implies, cf discussion leading to formulae (\ref{lala}) - (\ref{lalala}),
\begin{equation}
\label{hej}
\left(\cB(\cH) \otimes_{\pi} {\mathfrak T}\right)^* \supseteq \left(\cB(\cH) \otimes {\mathfrak K}\right)^* \supseteq \left(\cB(\cH) \otimes_{\epsilon} {\mathfrak K}\right)^*
\end{equation}
But $\cB(\cH) \otimes {\mathfrak K}$ is a $C^*$ algebra which is also an algebra of operators
acting on a specific Hilbert space. Therefore, the concept of normal (as well as singular) functionals on this $C^*$algebra is a meaningful one (cf. Section 10.1 in Kadison-Ringrose book \cite{KR}).
One can say even more. Namely, as the projective norm $\| \cdot \|_{\pi}$ is submultiplicative, the involution $^*$ is isometric for this norm then $\cB(\cH) \otimes_{\pi} \mathfrak T$ is a $*$-Banach algebra acting, again, on the specific Hilbert space. And this Banach $^*$-algebra is a subspace of the $C^*$-algebra $\cB(\cH) \otimes \mathfrak K$. Hence, any functional on $\cB(\cH) \otimes \mathfrak K$ leads to a well defined functional on $\cB(\cH) \otimes_{\pi} \mathfrak T$.
Further, note that any bounded linear functional $\varrho$ on $\cB(\cH) \otimes \mathfrak K$
can be uniquely expressed in the form $\varrho_1 + \varrho_2$ where $\varrho_1$ is a ultraweakly continuous while $\varrho_2$ is a singular one. For more details on a characterization of normal functionals we refer the reader to Section 10.1 in \cite{KR}).
Therefore
\begin{corollary}
There is a possibility to select normal functionals $\{ \psi \}$ on $\cB(\cH) \otimes_{\pi} {\mathfrak T}$.
Thus, one can select a family of positive maps having ``bp density matrices''. In particular, 
one can generalize the definition of ${\mathfrak D}_0$ for the general case and an analysis of its exposed points is also possible. Consequently, a generalization of a characterization of positive unital maps in terms of ``generalized'' Choi matrices is possible.
\end{corollary}

However, it is worth pointing out that there is no hope to get in this way a characterization of all positive unital maps. Namely, $^*$-weak continuity selects normal maps and only such maps are ``good'' candidates for maps associated with ``generalized'' Choi matrices. 
In other words, in Lemma \ref{pierwszylemat}, weak topologies should be taken into account.
Let us elaborate this point. We fix a positive unital map $\phi$ in $\cL(\cB(\cH))$. Then, Lemma \ref{pierwszylemat} gives the corresponding functional $\tilde{\phi}$ on $\cB(\cH) \otimes_{\pi} \cB(\cH)_*$ and 
the corresponding bilinear  form $B_{\phi}$. To take into account that $\phi$ is also normal we should complete our analysis with questions concerning weak topologies. To this end let us
consider $(\cB(\cH), \sigma(\cB(\cH),\cB(\cH)_*) \equiv \cB(\cH)_{\sigma}$ i.e. the set $\cB(\cH)$ equipped with $\sigma$-weak-$^*$ topology, and similarly $({\mathfrak K}, \sigma({\mathfrak K}, {\mathfrak T}))\equiv {\mathfrak K}_{\sigma}$, $({\mathfrak T}, \sigma({\mathfrak T}, \cB(\cH))\equiv {\mathfrak T}_{\sigma}$ .
Obviously we are using the dual pairs 
\begin{equation}
<{\mathfrak K}, \mathfrak T>, \qquad < {\mathfrak T}, \cB(\cH)>
\end{equation}
with the following identifications:
\begin{itemize}
\item ${\mathfrak T} \equiv \cB(\cH)_*$
\item ${\mathfrak K}^* \equiv \mathfrak T$
\item ${\mathfrak T}^* \equiv \cB(\cH)$
\end{itemize}

\textit{The crucial point to note here is that formulae (\ref{lala}) and (\ref{lala1})
can be generalized for the projective tensor product of locally convex spaces $\cB(\cH)_{\sigma}$ and ${\mathfrak T}_{\sigma}$,}
(see \cite{Jarch} for details).  
In other words, we will consider two locally convex vector spaces $\cB(\cH)_{\sigma}$ and ${\mathfrak T}_{\sigma}$ and we will be concerned with the projective tensor topology on 
$\cB(\cH)_{\sigma} \odot {\mathfrak T}_{\sigma}$.

To get a ``weak topology'' counterpart  of (\ref{lala}) one has, (see Proposition 2, page 329 in \cite{Jarch}):
\begin{proposition}
Let $B: \cB(\cH)_{\sigma} \times {\mathfrak T}_{\sigma} \to \bC$ be a bilinear form. Then $B$ is separately continuous (with respect to weak topologies) if and only if the linear map $T_B: x \mapsto B(x, \cdot)$ is continuous (again with respect to weak topologies).
\end{proposition}

On the other hand, see Section 15.1 in \cite{Jarch} and Section 2 of the paper, there is also one-to-one correspondence $\Theta$ between the dual space $(\cB(\cH)_{\sigma} \otimes_{\pi} {\mathfrak T}_{\sigma})^*$ and continuous bilinear forms on $\cB(\cH)_{\sigma} \times {\mathfrak T}_{\sigma}$.
Here $\cB(\cH)_{\sigma} \otimes_{\pi} {\mathfrak T}_{\sigma}$ stands for $\cB(\cH)_{\sigma} \odot {\mathfrak T}_{\sigma}$ equipped with the projective topology originated from the corresponding weak topologies.
Denote by $B(\cB(\cH)_{\sigma}, {\mathfrak T}_{\sigma})$ the set of all continuous bilinear forms on $\cB(\cH)_{\sigma} \times {\mathfrak T}_{\sigma}$. Then
$\Theta: (\cB(\cH)_{\sigma} \otimes_{\pi} {\mathfrak T}_{\sigma})^* \to B(\cB(\cH)_{\sigma}, {\mathfrak T}_{\sigma})$
is given by $T \to T\circ\otimes$. Note that the tensor product $\otimes$ is continuous. 
Let us apply the above results to $\phi \in \cL(\cB(\cH))$.
Since $\phi$ is a normal map,  the corresponding bilinear form $B_{\phi}$ is also separately weakly continuous. But this implies that $\tilde{\phi}$ also has this property.
 Furthermore, as the weak topology $\sigma({\mathfrak T}, \cB(\cH))$ is finer than the weak (operator) topology restricted to $\mathfrak T$, $\tilde{\phi}$
can be continuously extended to weakly continuous functional on $\cB(\cH) \odot \cB(\cH)$ and this gives rise to a ``generalized'' Choi matrix. Thus we got:
\begin{corollary}
A version of Lemma \ref{pierwszylemat} for $\cB(\cH)_{\sigma}$ and ${\mathfrak T}_{\sigma}$ offers a generalization of Choi-Jamiolkowski correspondence for infinite dimensional case.
\end{corollary}

We close this Section with a remark concerning inclusions (\ref{hej}). Namely,
it is worth noting that we are studying the subset of maps which is large enough to contain integral maps corresponding integral functionals.

\section{Structure of positive maps versus PPT states}

As the second example of applications of the Grothendieck approach we present a characterization of PPT states (see \cite{MMO}). Again we will exploit 
links between tensor products and certain subsets of states.
We wish to emphasize that such the approach is necessary if one wish to consider quantum systems (so infinite dimensional; cf arguments following Theorem 7).
As PPT states are ``dual'' to decomposable maps (see \cite{I}, and Section 1) we need an adaptation of the Basic Lemma for CP and co-CP maps. In \cite{MMO} we got:
\begin{lemma}
\label{drugi lemat}
(1) Let
$\mathcal{B}\left[ \ \mathcal{B}\left( \mathcal{H}\right) ,\mathcal{B(K)}%
_{\ast }\right] $ stand for the set of all linear, bounded, normal (so weakly$^*$-continuous) maps from $\cB(\cH)$ into $\cB(\cH)_{\ast}$.
There is an isomorphism $\psi \longmapsto \tilde{\psi} $ between $%
\mathcal{B}\left[ \ \mathcal{B}\left( \mathcal{H}\right) ,\mathcal{B(K)}%
_{\ast }\right] $ and $\left( \mathcal{B(H)\otimes B(K)}\right)_{\ast }$
given by 
\begin{equation}
\tilde{\psi} \left( \sum_{i}a_{i}\otimes b_{i}\right) =\sum_{i}Tr_{\mathcal{K}}\psi
\left( a_{i}\right) b_{i}^{t},\text{\ \ }a_{i}\in \mathcal{B}\left( \mathcal{%
H}\right) ,\text{ }b_{i}\in \mathcal{B}\left( \mathcal{K}\right) .
\end{equation}%
The isomorphism is isometric if $\tilde{\psi}$ is considered on $\mathcal{B(H){\otimes}_{\pi} B(K)}$. 
Furthermore $\tilde{\psi} $ is positive on $(\cB(\cH)\otimes \cB(\cK))^+$ if and only if $\psi $ is complete positive.

(2) There is an isomorphism $\phi \longmapsto \tilde{\phi} $ between $%
\mathcal{B}\left[ \ \mathcal{B}\left( \mathcal{H}\right) ,\mathcal{B(K)}%
_{\ast }\right] $ and $\left( \mathcal{B(H)\otimes B(K)}\right)_{\ast }$
given by 
\begin{equation}
\tilde{\phi} \left( \sum_{i}a_{i}\otimes b_{i}\right) =\sum_{i}Tr_{\mathcal{K}}\phi
\left( a_{i}\right) b_{i},\text{\ \ }a_{i}\in \mathcal{B}\left( \mathcal{H}%
\right) ,\text{ }b_{i}\in \mathcal{B}\left( \mathcal{K}\right) .
\end{equation}%
The isomorphism is isometric if $\tilde{\phi}$ is considered on $\mathcal{B(H){\otimes}_{\pi} B(K)}$. 
Furthermore $\tilde{\phi} $ is positive on $(\cB(\cH)\otimes \cB(\cK))^+$ iff $\phi $ is complete co-positive.
\end{lemma}

One of the consequences of this lemma is a possibility to defined so called {{\it entanglement mappings} \cite{BO}},  see also \cite{BO2} and \cite{MMO}. We recall that this concept is a starting point of the Belavkin-Ohya approach to a characterization of PPT states. Namely,
following this scheme, let us fix a normal state $\omega$ on $\cB(\cH) \otimes \cB(\cK)$, i.e. a density matrix $\varrho$ describing a composite state $\omega$ is fixed. Let $\sum_i \lambda_i |e_i><e_i|$ be its spectral decomposition.  We will need
a linear bounded
operator $T_{\zeta }:\mathcal{K}\rightarrow \mathcal{H}\otimes \mathcal{K}$ defined 
by
\begin{equation}
T_{\zeta }\eta =\zeta \otimes \eta
\end{equation}
where $\zeta \in \mathcal{H}$, $\eta \in \mathcal{K}$.
Then, we are able to define the entanglement operator $H$
\begin{equation}
H:\mathcal{H}\rightarrow \mathcal{H}\otimes \mathcal{K}\otimes \mathcal{K}
\end{equation}%
by the formula:
\begin{equation}
H\zeta =\underset{i}{\sum }\lambda _{i}^{\frac{1}{2}}\left( J_{\mathcal{H}%
\otimes \mathcal{K}}\otimes T_{J_{\mathcal{H}}\zeta }^{\ast }\right)
e_{i}\otimes e_{i}
\end{equation}%
where $J_{\mathcal{H}\otimes \mathcal{K}}$ is a complex conjugation defined
by $J_{\mathcal{H}\otimes \mathcal{K}}f\equiv J_{\mathcal{H}\otimes \mathcal{%
K}}\underset{i}({\sum }\left( e_{i}^{\cdot },f\right) e_{i}^{\cdot })=\underset%
{i}{\sum }\overline{\left( e_{i}^{\cdot },f\right) }e_{i}^{\cdot }$ where $%
\left\{ e_{i}^{\cdot }\right\} $ is any CONS (complete orthonormal system) extending (if necessary) the
orthogonal system $\left\{ e_{i}\right\} $ determined by the spectral
resolution of $\rho $. ( $J_{\mathcal{H}}$ is defined analogously with the
spectral resolution given by $H^{\ast }H$; using the explicit form of $H$, easy calculations show that the spectrum of $H^*H$ is discrete.)
Having the entanglement operator $H$, we can define the entangling mapping $\phi$, which is the crucial ingredient of B-O approach. Namely:
\begin{equation}
\label{la}
\phi \left( b\right) =\left( H^{\ast }\left( 1\otimes b\right) H\right)
^{t}=J_{\mathcal{H}}H^{\ast }\left( 1\otimes b\right) ^{\ast }HJ_{\mathcal{H}.%
}
\end{equation}%:

The properties of the entangling mapping and its dual  are collected in the next proposition which was proved in \cite{MMO}.

\begin{proposition}
\label{PPT}
The entanglement mapping

(i) $\phi ^{\ast }:\mathcal{B}\left( \mathcal{H}\right) \rightarrow \mathcal{%
B}\left( \mathcal{K}\right) _{\ast }$ has the following explicit form%
\begin{equation}
\phi ^{\ast }\left( a\right) =Tr_{\mathcal{H}\otimes \mathcal{K}%
}Ha^{t}H^{\ast }
\end{equation}

(ii) The state $\omega $ on $\mathcal{B}\left( \mathcal{H}\otimes \mathcal{K}%
\right) $ can be written as 
\begin{equation}
\label{17}
\omega \left( a\otimes b\right) =Tr_{\mathcal{H}}a\phi \left( b\right) =Tr_{%
\mathcal{K}}b\phi ^{\ast }\left( a\right)
\end{equation}
where $\phi$ was defined in (\ref{la}).
\end{proposition}

Now, turning to the characterization of PPT states we note that, by definition, we note that any such state composed with partial transposition is again a state. This observation and  Proposition (\ref{PPT}) leads to (see \cite{MMO})
\begin{corollary}
PPT states are completely characterized by entanglement mappings $\phi^{\ast}$ which are both CP and co-CP.
\end{corollary}

To sum up these notes we arrived at
\begin{conclusion}
The Grothendieck approach to tensor products of locally convex spaces provides a useful tools for an analysis of crucial problems of the theory of quantum maps and states. 
In particular, it offers a possibility to generalize Choi matrices for genuine quantum systems. Consequently, also, it is possible to generalize the description of positive maps given in \cite{JMP}.
\end{conclusion}

\section{Acknowledgments}

The author would like to thank E. St{\o}rmer for bringing Holevo's work to his attention. He also thanks
the support of the grant of the Foundation for Polish Science TEAM project cofinanced by EU European Regional Development Fund.

\end{document}